\documentclass[11pt,a4paper]{article}

\usepackage{amssymb}
\usepackage{graphicx}

\newcommand{\be}{\begin{equation}}
\newcommand{\ee}{\end{equation}} 
\newcommand{\bea}{\be \begin{array}{rcl}}
\newcommand{\eea}{\end{array}\ee}
\newcommand{\ba}{\begin{array}}
\newcommand{\ea}{\end{array}}
\newcommand{\ra}{\rightarrow}

\newcommand{\Sp}{\;\;\;\;}
\newcommand{\av}[1]{\langle #1 \rangle}
\renewcommand{\Re}{{\rm Re}}

\newcommand{\captionW}[1]{\hspace*{1.3cm}\parbox{13.9cm}{\caption{\footnotesize #1}} }

\newcommand{\bra}[1]{\langle{#1}|}
\newcommand{\ket}[1]{|{#1}\rangle}

\newcommand{\hU}{\hat{U}}
\newcommand{\hH}{\hat{H}}

\newcommand{\vb}{\overline{v}}

\newcommand{\Tt}{\tilde{T}}

\newcommand{\dl}{\delta}
\newcommand{\eps}{\epsilon}

\newcommand{\lm}{\lambda}

\newcommand{\bw}{}

\title{Quantum Equilibrium in Stochastic  de Broglie-Bohm-Bell Quantum Mechanics}
\author{Jeroen C. Vink  \thanks{\parbox{13.5cm}{
                            Shell Global Solutions International B.V.,
                               Gasweg 31, 1031 HW Amsterdam, The Netherlands.
		      Email: Jeroen.Vink@Shell.com} } }

\begin{document}


\maketitle

\abstract{ This paper investigates dynamical relaxation to quantum equilibrium in the stochastic
              de Broglie-Bohm-Bell formulation of quantum mechanics. The time-dependent probability distributions
are computed as in a Markov process with slowly varying transition matrices.
Numerical simulations, supported by exact results for the large-time behavior of sequences of (slowly
varying) transition matrices, confirm previous findings that indicate that de Broglie-Bohm-Bell dynamics 
allows an arbitrary initial probability distribution to relax to quantum equilibrium; i.e., there is no
need to  make the ad-hoc assumption that the initial distribution of particle locations has to be identical to the 
initial probability distribution prescribed by the system's initial  wave function. 
The results presented in this paper moreover suggest that the intrinsically stochastic nature of  
Bell's formulation, which is arguable most naturally formulated on an underlying discrete space-time, is sufficient 
to ensure dynamical relaxation to quantum equilibrium for a large class of quantum systems without the need 
to introduce coarse-graining or any other modification in the formulation.
}


\section{Introduction} \label{sect1}
The formulation of quantum mechanics due to de Broglie, Bohm and Bell (dBBB) has many advantages over the
conventional Copenhagen interpretation (cf.~for example refs.~\cite{BohmHiley93,Goldstein21,Barrett19}).
These advantages and benefits hold for the original, causal formulation originally proposed by de Broglie
\cite{deBroglie56} and rediscovered and perfected by Bohm \cite{Bohm52} as well as for the stochastic 
version introduced by Bell \cite{BellCh19} and further developed in refs.~\cite{Vink93,Vink18,Vink22}.
The benefits and attractive features notwithstanding, these Bohm-type formulations continue to be criticized.
One of the more fundamental challenges concerns a seemingly ad-hoc assumption that is required to
reconcile the computed probabilities in  dBBB formulations with the established probability interpretation
and experimental results of quantum mechanics.

In dBBB formulations, particles always have a well-defined position and move (either causally or stochastically) along
trajectories that are guided by the system's wave function. If the  probability distribution of the initial particle locations 
is the same as the probability distribution defined by the system's initial wave function, then the particle dynamics
is such that for all future times, the probability distribution of the particle locations is identical to the distribution 
defined by the time-dependent wave function. 
Recognizing this ad-hoc assumption on the initial particle distribution as a weakness in the formulation, already 
Bohm and Hiley proposed \cite{BohmHileyCh9} that it could be avoided if an arbitrary 
initial probability distributions would dynamically relax to the quantum ``equilibrium'' distribution. 
This was further explored and
substantiated in work by Valentini and others \cite{Valentini92,Valentini01, ValentiniWestman05,ColinStruyve10}
(see also ref.~\cite{DurrEtAl92} for an alternative approach to understand the role of probability in dBBB formulations).

Such a dynamical relaxation to quantum equilibrium is not at all obvious. In the causal Bohm formulation, the same dynamics that ensures that the probability distribution of the ensemble
of particles continues to reproduce the probabilities computed with the system's wave function, also implies
that any deviation from this distribution in the initial state will be preserved for all later times.
 Hence, the exact dynamics cannot accommodate a spontaneous relaxation to quantum equilibrium and one
has to invoke a coarse-graining procedure where particle locations are evaluated as averaged values, for example
on a spatial lattice.
Owing to underlying chaotic particle dynamics, these coarse grained locations could be, and in 2D examples studied in 
refs.~\cite{Valentini01,ValentiniWestman05,ColinStruyve10} in fact are, found to relax to quantum equilibrium -
with a relaxation time that Valentini argues \cite{Valentini92,Valentini01} scales as
\be
   t^{eq} \propto m\hbar^2/\varepsilon (\Delta P)^{3}.   \label{Eq_teq}
\ee
Here, $m$ is the particle mass, $\varepsilon$ the coarsening scale and $\Delta P$ the momentum spread
in the system's wave function.

Unlike the causal Bohm formulation, Bell's formulation is stochastic, and in the version of refs.~\cite{Vink93,Vink22} 
space is discrete with particle locations limited to the sites of a spatial lattice.
Here, the dynamics does not imply that an initial non-equilibrium must persist - in fact, the similarity
of this stochastic formulation to a Markov process might  suggest that equilibration could happen without
the need for further modifications of the dynamics.
This is explored in the present paper  
using numerical simulations of a 2D quantum system describing a free particle in a box.
In these simulations there is no need to compute large ensembles of  trajectories from which the
probability distribution of the particle location must be determined. 
Provided the systems are reasonably sized, the time evolution of the
probability distributions can be computed directly, applying the time-dependent transition matrices of the system 
to the evolving probability distribution.
The definition of these transition matrices and their features, along with the similarities
 and difference between a Markov and a Bell process  will be further discussed in section \ref{sect2} below.

The results of these numerical simulations show
rapid and robust relaxation to quantum equilibrium, which by the nature of the (weak) ergodicity of the process
automatically applies to all (non-equilibrium) initial probability distributions.
These simulations can be performed for various values of the particle mass, lattice distance and momentum
spread in the wave functions, with results for the equilibrium time  that suggest a slightly different
 scaling behavior than proposed in (\ref{Eq_teq}):
\be
  t^{eq} \propto  mL\hbar / a (\Delta P)^2.  \label{Eq_teq2}
\ee
Here, $a$ is the lattice distance and $L$ the box dimension. As in Eq.~(\ref{Eq_teq}) this time scale is linear
in the particle mass and linear in the inverse lattice distance - which acts as a natural substitute for
the coarsening scale required in the causal dynamics. 
The dependence on momentum spread is less severe, with one (inverse) factor  replaced by a factor $L$.

These results confirm that dBBB formulations do not need the additional assumption that the initial distribution
of particle locations has to be identical to the initial probability distribution prescribed by the system's initial 
wave function. They moreover suggest that the intrinsically stochastic nature of  Bell's formulation, which arguable is
most naturally formulated on an underling discrete space-time, is sufficient to ensure dynamical relaxation
to quantum equilibrium for a large class of quantum systems - without the need to introduce coarse-graining
or any other modification in the formulation.

The remainder of the paper is organized as follows. The next section starts with a review of the discrete 
space-time version of the dBBB formulation \cite{Vink22}, followed by a brief summary of Markov processes 
and their convergence, after which the discussion is extended to explore properties of the Bell-process
in which the Markovian transition matrices are time-dependent. Section \ref{sect3} begins with the
description of the 2D quantum systems, followed by an overview of the simulation results for the
lattice distance, particle mass and momentum spread dependent equilibrium times and, finally, 
section \ref{sect4} contains a summary with concluding remarks.

%
%

\section{Markov Processes with Time-dependent Transition Probabilities}  \label{sect2}
\subsection{Stochastic Bell Dynamics} \label{sect2sub1}
Bell's stochastic version \cite{BellCh19} of Bohm mechanics in the  discrete time version of 
ref.~\cite{Vink22} can be formulated as follows.
The system's quantum state evolves according to  Schr\"odinger dynamics as
\be
   \ket{\psi(t+\eps)} = \hU\ket{\psi(t)},   \label{Eq_SE}
\ee
where $t$ is the discrete time\footnote{As was briefly discussed in ref.~\cite{Vink22}
the time step size $\eps$ could be time (index) dependent, with a magnitude that is self-consistently determined by
the system's dynamics. Here, such a time dependence will be ignored, as it does not materially impact
the results presented below.} that progresses in steps $\eps$ and
\be
   \hU=e^{-i\eps \hH}  \label{Eq_U}
\ee
is the evolution operator with its usual dependence on the system's (time-independent) Hamiltonian $\hH$. 
Also space is discrete and finite such that particle configurations
can be labeled with indices $n=1,\dots,N$; in such a location representation, Eq.~(\ref{Eq_SE}) takes the form
\be
   \psi_n(t+\eps)= \sum_m U_{nm}\psi_m(t),   \label{Eq_Psit}
\ee
with $\psi_n(t) = \bra{n}\psi(t)\rangle$ and $U_{nm}=\bra{n}\hU\ket{m}$.
As was shown in ref.~\cite{Vink22}, the time-dependent configuration probabilities obey a discrete-time continuity
equation,
\be
   P_n(t+\eps) = P_n(t) + \sum_m J_{nm}(t),  \label{Eq_PJ}
\ee
with 
\be
    P_n(t)=\vert\psi_n(t)\vert^2   \label{Eq_Pt}
\ee
and 
\be
   J_{nm}(t) = \Re(\psi^{*}_n(t+\eps)U_{nm}\psi_m(t))  - (n \leftrightarrow m). \label{Eq_Jt}
\ee
Instead of using the recursion defined in Eq.~(\ref{Eq_PJ}), the probability distribution $P_n(t)$, which will
also be referred to as ``quantum equilibrium distribution'', at time  $t=k\eps$ given an initial state $\psi(0)$ 
can  equally well be directly computed from the Schr\"odinger equation (\ref{Eq_SE}) as
\be
   P_n(t) = \vert \sum_m(U^k)_{nm}\psi_m(0)\vert^2.   \label{Eq_PSE}
\ee

These time-dependent probabilities can also be generated from an ensemble of stochastically evolving configuration
trajectories $\{n_i(t)\}_{i=1}^M$. Here, $n(t)$ is the index for the particle configuration at time $t$, and
$i$ labels the configuration trajectories in the ensemble of size $M$. 
The trajectories can be generated using transition probabilities $T_{nm}(t)$ defined as \cite{BellCh19,Vink93}
\be
   T_{nm}(t) = \theta(J_{nm}(t))J_{nm}(t)/P_m(t),    \label{Eq_Tnm}
\ee
\be
   T_{mm}(t) = 1 - \sum_{n\ne m} T_{nm}(t),   \label{Eq_Tmm}
\ee
where $\theta(x)$ is the Heaviside step function and $T_{nm}(t)$ is the probability that configuration 
$m=n_i(t)$ at time step $t$ changes to configuration $n=n_i(t+\eps)$ at time  $t+\eps$.
The configuration probabilities can then be computed from the ensemble as
\be
P_m(t) = \lim_{M\ra\infty} \sum_{i=1}^M \dl_{m,n_i(t)}/M.     \label{Eq_Ptraj}
\ee

Equivalently, the time dependence of the configuration probabilities (\ref{Eq_Ptraj}) can be computed
using the following time dependent Master Equation,
\be
   P_n(t+\eps) = P_n(t) + \sum_m\left( T_{nm}(t)P_m(t) - T_{mn}(t)P_n(t)\right).  \label{Eq_ME}
\ee
Since the transition matrices $T(t)$ are stochastic matrices,
\be
   T_{nm}(t) \ge 0, \forall n,m,t; \Sp \sum_n T_{nm}(t) = 1, \forall m,t,  \label{Eq_Tstoch}
\ee
it follows that $\sum_m T_{mn}(t)P_n(t) = P_n(t)$ and hence Eq.~(\ref{Eq_ME}) can be rewritten in a 
more explicitly Markovian form as
\be
   P(t+\eps) = T(t)P(t).  
                          \label{Eq_Ptp1}
\ee
This in turn implies that the probability at time $t$ can be computed directly from the initial
probabilities as
\be
  P(t) = \Tt(t)P(0),    
                        \label{Eq_PtFromT}
\ee
where the cumulative transition matrix $\Tt(t)$ is defined as the backward product of the preceding transition 
matrices $T(t)$,
\be
   \Tt(t)\equiv   T(t-\eps)T(t-2\eps)\dots T(0).   \label{Eq_Tt}
\ee
As was shown in ref.~\cite{Vink22}, $P(t)$ computed from Eq.~(\ref{Eq_PtFromT}) is identical to the quantum
distribution defined in Eq.~(\ref{Eq_PSE}) computed using the Schr\"odinger dynamics (\ref{Eq_SE})
 if the  distributions at time $0$ are the same, $P_m(0)=\vert\psi^*_m(0)\psi_m(0)\vert^2$, and provided the
time step size $\eps$ in Eq.~(\ref{Eq_U}) is sufficiently small to ensure that $T_{mm}(t)$ defined 
in Eq.~(\ref{Eq_Tmm}) is always non-negative.

\subsection{Markov Processes} \label{sect2sub2}
To set the stage for exploring if $P(t)$ defined in Eq.~(\ref{Eq_PtFromT}) also converges to the quantum
equilibrium distribution (\ref{Eq_PSE}) for arbitrary $P(0)$, it is worthwhile to review some properties of a 
(fixed) stochastic matrix $A$ and the associated Markov process,
\be
   \pi^{(k+1)} = A\pi^{(k)}.   \label{Eq_Markov}
\ee
Since the sum of the matrix elements in every column of $A$ is one, it  follows that $u^TA = u^T$ for a unit-elements 
vector $ u=(1,1,\dots,1)^T$  and hence $u$ is a left eigenvector of $A$ with eigenvalue $1$.
 The Gershgorin circle theorem (see e.g.~\cite{GolubVanLoan83}) applied to the rows of the transposed 
matrix $A^T$ then implies that all other eigenvalues have norm less or equal to one.
If $A$ is also primitive, i.e., for a sufficiently large power $k$ all matrix elements of $A^k$ are positive,
 then the Perron-Frobenius theorem (see, e.g., \cite{Sternberg96,Seneta81}) states that the eigenvalue 
$\lm^0=1$ is unique and strictly larger than all other eigenvalues $\lm^s$ (i.e., $\vert\lm^s\vert < 1-\delta$ 
with $\delta>0$ and $s>0$).  
If this is the case, only the dominant eigenmode remains in the 
spectral decomposition of $A^k$ for sufficiently large  $k$:
\be
   \lim_{k\ra\infty} A^k = \pi^{eq}u^T,   \label{Eq_Meq}
\ee
where $\pi^{eq}$ is the probability-normed right eigenvector with eigenvalue $1$ associated with $u$,
\be
   A\pi^{eq} = \pi^{eq}, \Sp u^T\pi^{eq} = 1.
\ee
This implies that
the Markov process with such a transition matrix converges to the unique equilibrium distribution $\pi^{eq}$, 
\be
  \pi^{eq}= \lim_{k\ra\infty} A^k \pi^{(0)},    \label{Eq_pieq}
\ee
for any initial probability distribution $\pi^{(0)}$.

\subsection{Markov Process with Time-Dependent Transition Matrix} \label{sect2sub3}
The Markov process (\ref{Eq_Markov}) resembles the Bell process (\ref{Eq_Ptp1}),
but besides the formal similarity, there is the important difference that the transition matrices in
(\ref{Eq_Ptp1}) are  time dependent.  It is therefore not clear how much
one can learn from the spectrum of individual  matrices. 
However, since every $T(t)$ is a stochastic
matrix,  it remains the cases that for all $t$, $T(t)$ has (at least) one eigenvalue $1$ with left eigenvector 
$u = (1,\dots,1)^T$ and all other eigenvalues have an absolute value less than or equal to $1$.
It also remains the case that the  properties of the spectral decomposition of the cumulative transition matrix,
\be
   \Tt(t)= \sum_{s=0}^{N-1}  \lm^s(t) v^s(t) \vb^{sT}(t),         \label{Eq_OneV}
\ee
determine if
the Bell process (\ref{Eq_Ptp1}) will equilibrate to a late-time behavior that is independent of the initial state.
Here, $v^s(t)$ and $\vb^s(t)$ are the right and left eigenvectors of $\Tt(t)$, with eigenvalue $\lm^s(t)$.
This equilibration will be established if the process is ``weakly ergodic''\footnote{See ref.~\cite{Seneta81} for the
formal definition of weak (and strong) ergodicity.}, i.e., only a single  eigenmode with 
eigenvalue $1$ dominates  in the cumulative transition matrix for times (much) larger than an equilibrium 
time $t^{eq}$:
\be
     \Tt(t)  \ra \lm^0(t) v^0(t) \vb^{0T}(t)  \equiv \pi(t)u^T \Sp {\rm for} \Sp t \gg t^{eq},                            \label{Eq_Teq}
\ee
with $\vb^0(t)=u$ the constant left eigenvector with eigenvalue  $\lm^0=1$ and $v^0(t)=\pi(t)$ the corresponding 
time-dependent right eigenvector.
If this is the case, this right eigenvector $\pi(t)$ necessarily will be equal to the quantum distribution $P(t)$,
because $\Tt(t)\pi(0)=\pi(t)$ for any initial distribution $\pi(0)$ and by construction $\Tt$ produces
the quantum equilibrium distribution $P(t)$ if the initial distribution $\pi(0)=P(0)$.

This shows that it is sufficient that $\Tt(t)$ satisfies the large time behavior (\ref{Eq_Teq})
to obtain a Bell process (\ref{Eq_Tt}) that automatically converges to reproduce quantum dynamics
irrespective of the initial probability distribution. 
The late-time behavior (\ref{Eq_Teq}) is also necessary, since the contribution from  other modes,
i.e., terms $\lm^s v^s(t)(\vb^{sT}(t)\pi(0))$ with $s>0$, can only be independent of $\pi(0)$
if $(\vb^{sT}(t)\pi(0))$ is the same for arbitrary  $\pi(0)$, which can only be the 
case\footnote{More precisely (dropping the argument $(t)$):  Suppose $\Tt\pi$ is invariant under an 
infinitesimal shift of $\pi$ of the form,  $\pi \ra \pi + \dl^j - \dl^{j'}$ with $\dl^j$ a vector which is zero everywhere, 
except on location $j$ where it has value $\eps$. Then the infinitesimal change of $\Tt$ must be zero for all pairs
$j,j'$, i.e., $\dl(\sum_m\Tt_{nm}\pi_m) = \sum_m\Tt_{nm}(\dl^j_m - \dl^{j'}_m) = 
\eps \sum_s \lm^s v^s_n(\vb^s_j - \vb^s_{j'})    = \eps(\Tt_{nj} - \Tt_{nj'}) = 0$, 
which shows that $\Tt$ must have identical columns, as in Eq.~(\ref{Eq_Teq}). }
 if $\vb^s(t)\propto u$.  
However, there cannot be additional left eigenvectors proportional to $u$, since they would have to be
orthogonal to the right eigenvector $v^0(t)$.


To establish the result (\ref{Eq_Teq}), it  is therefore sufficient to show that the Markov process generating the
cumulative transition matrix $\Tt(t)$ is weakly ergodic. 
 In practical terms, it is then sufficient to show that
 for large $t$ the next-largest eigenvalue  $\lm^1(t)$ (and hence every subsequent eigenvalue) 
approaches zero, $\vert\lm^1(t) \vert \ra 0$ for $t\ra \infty$. 
Slightly more specifically and in analogy with the $k$ dependence of the
eigenvalue spectrum of $A^k$ in a Markov process (\ref{Eq_Meq}), it is sufficient that the time dependence 
of the eigenvalues can be expressed as,
\be
 \vert\lm^s(t)\vert \approx e^{-c(t) w_s},   \label{Eq_Lm}
 \ee
with $c(t)$ a positive (real) value that increases with $t$ and $w_s$ a positive weight factor that increase
with increasing eigenvalue index $s$ and $w_0=0$. 
Somewhat remarkably, the results of the numerical experiments discussed below can be described with 
weights $w_s \approx s$ and coefficients $c(t)$  that grow (approximately) linearly with time,
$ c(t) \approx  c_0 t$.
The coefficient $c_0$ then can be taken as a measure for the inverse equilibrium time, $c_0 \approx 1/t^{eq}$.

Before proceeding to the next section, which discusses numerical evaluations of the eigenvalue spectra of $\Tt(t)$, 
it is worthwhile to briefly contemplate how much one could learn from the properties of the individual matrices $T(t)$.

Since for a normal Markov process the matrices $A$ must be primitive, 
it may be reasonable to assume that  also the $T(t)$ matrices (at least predominantly) must be primitive, 
or equivalently, have a single eigenvalue $1$ and sub-leading eigenvalues that are well-separated from $1$.
Since matrices $T(t+\eps)$ and $T(t)$ are ``almost the same'' (because the time step size $\eps$ is very small),
it is then not unreasonable to assume that subdominant eigenmodes will be suppressed in each iteration
 by a factor roughly equal to the modulus of their eigenvalue $\vert\lm^s(t)\vert$.
Interestingly,  sequences of slowly varying non-negative primitive matrices have been studied 
in ref.~\cite{Artzrouni91}. There, Artzrouni proves that
for a sequence of slowly varying non-negative primitive matrices\footnote{Technically, the matrices must furthermore
be bounded and have non-zero elements well-separated from zero.} $T(t)$ with
 $\|T(t+\eps)-T(t)\| < \varepsilon_0$, there is an $\varepsilon_0>0$ such that
 the backward product  $\Tt(t)$ is weakly ergodic. Weak egodicity in this more generic
 situation implies that for all $t$ the product matrix $\Tt(t)$ can be written as,
\be
   \Tt_{nm}(t) = (w_m+\varepsilon_{nm}(t))L_n(t),  \label{Eq_wErg}
\ee
with constant, positive weight vector  $w$ and time-dependent, positive vector $L(t)$
and  $\varepsilon_{nm}(t)\ra 0$ for $t\ra \infty$.
For matrices that
are furthermore stochastic, the columns of $\Tt(t)$ always sum up to one and hence
the weights must be equal, $w_m=w$, such that $wL(t)$ becomes the dominant right
eigenvector of $\Tt(t)$,  $wL(t)=v^0(t)$.


This  suggests that under suitable, not necesserily very stringent conditions, the Bell process will obtain 
weak ergodicity (i.e., the cumulative transition matrix $\Tt(t)$ will assume the form Eq.~(\ref{Eq_Teq})) 
and  hence it will support the desired relaxation to quantum equilibrium for arbitrary initial probability distributions. 
Unfortunately, it is not straightforward to upfront determine if these  conditions (primitivity, sufficiently slowly varying)
apply to matrices $T(t)$ associated with a specific quantum system.
Therefore numerical simulations will still be needed to further explore and quantify the large-time 
behavior of $\Tt(t)$.

%
%

\begin{figure}[tth]
\begin{centering}
\scalebox{0.8}{\includegraphics[height=8.cm]{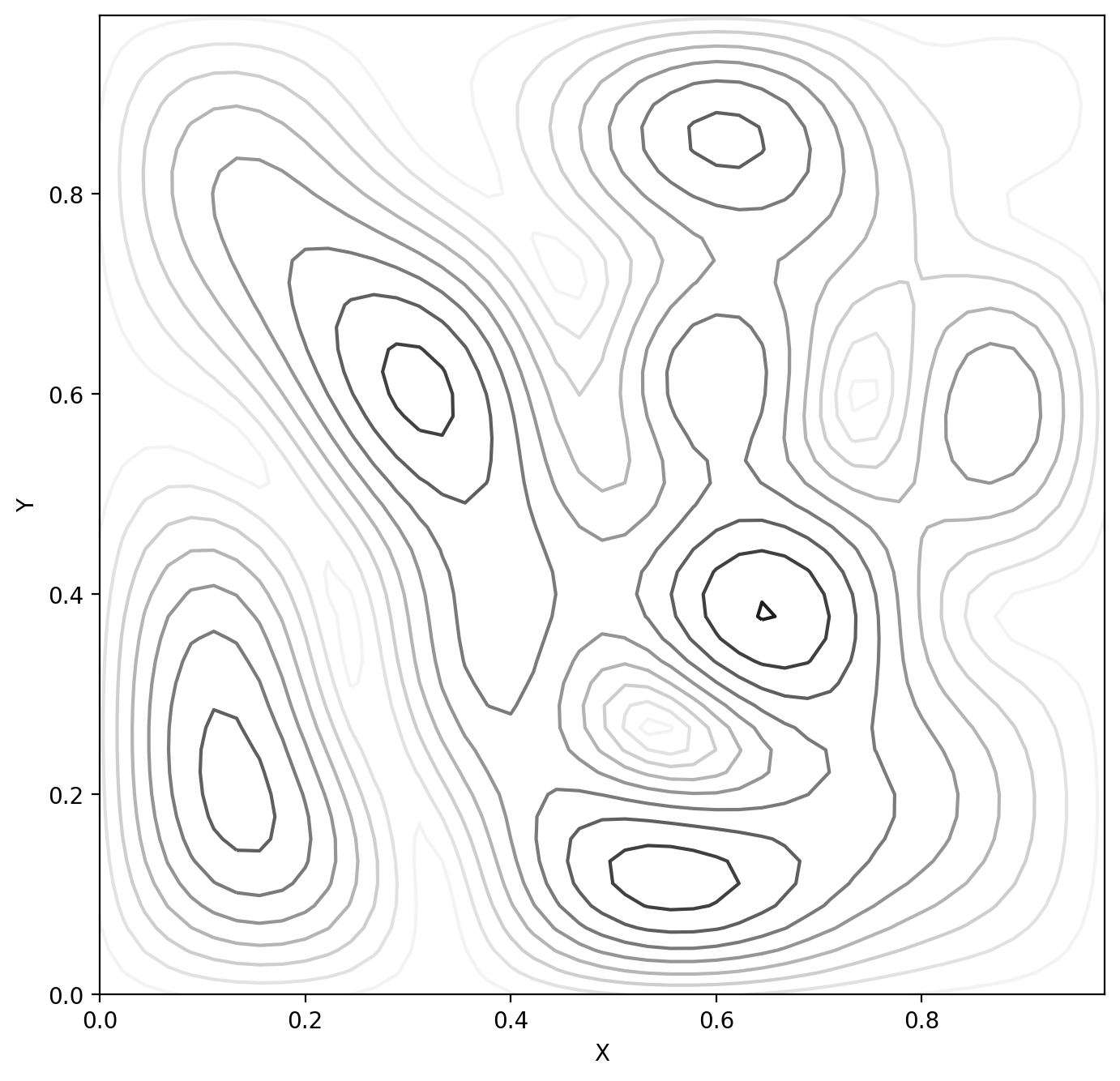}}
\scalebox{0.8}{\includegraphics[height=8.cm]{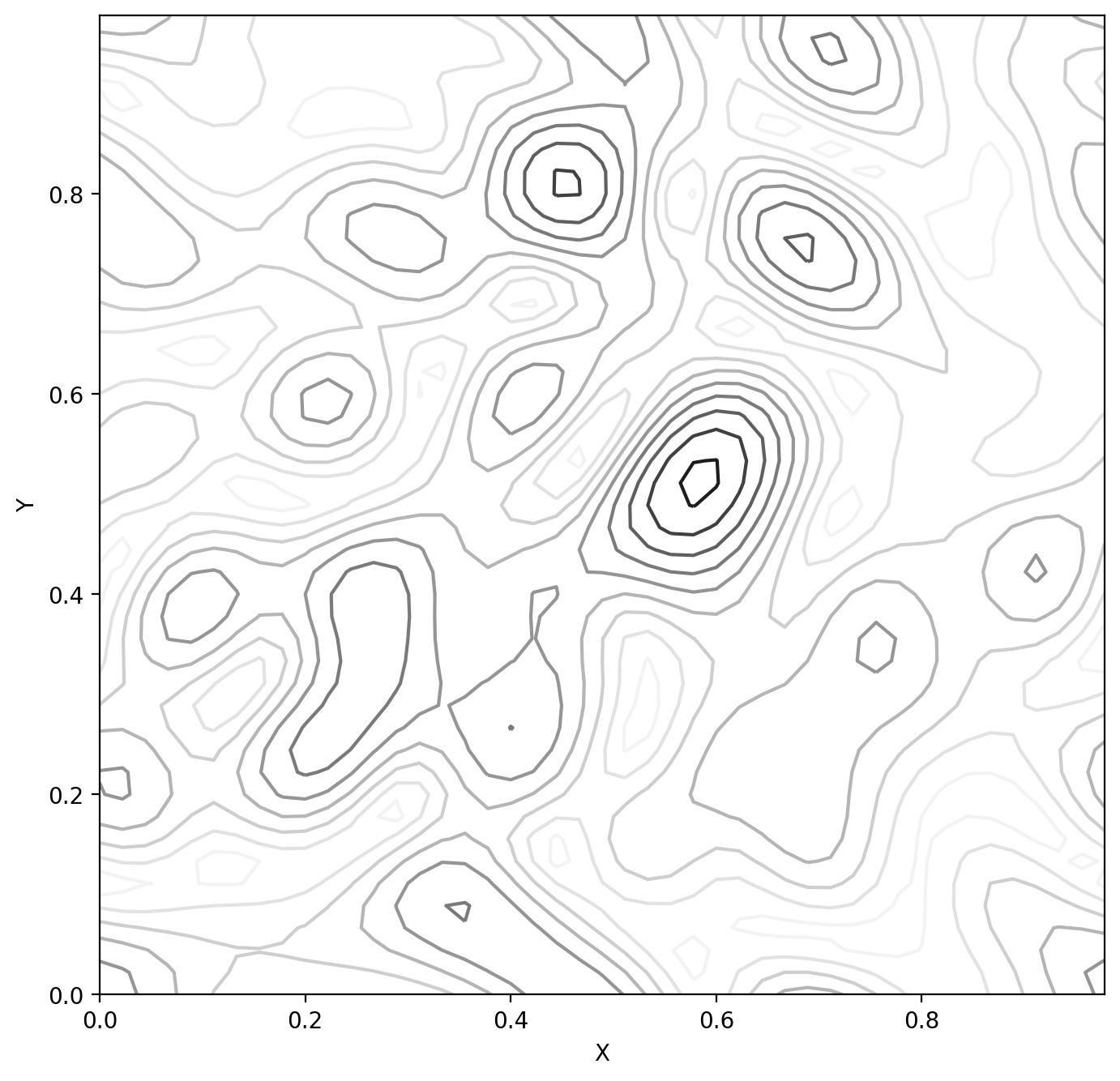}}

\end{centering}

\captionW{Contour maps of an initial probability distribution with low $\Delta P$, $N_k=4$ (left) and high 
$\Delta P$, $N_k=7$ (right), displayed in the 2D plane. 
The box dimension $L$ is set to 1 and is discretized with 45 lattice sites in each direction.
\label{Fig_1}}
\end{figure}

\section{Numerical Experiments} \label{sect3}
\subsection{Model Definition} \label{sect3sub1}
As in refs.~\cite{ValentiniWestman05,ColinStruyve10} the quantum systems investigated in the present
paper describe a free particle in a 2D box. Unlike in this previous work, the space within the box 
 is a square lattice of size $L\times L$ with $N$ lattice sites in each
direction.  The wave functions $\psi_x$ on this lattice  have Dirichlet boundary conditions $\psi=0$,
as in a box with impenetrable walls. In order to assess the impact of  boundary conditions, also periodic boundary conditions will be briefly considered.

The Hamiltonian for a free particle on this 2D lattice takes the form,\footnote{  
Units are such that $\hbar = c = 1$ and the  scale is set using the linear box dimension $L$. }
\be
   H_{x,y} = \frac{1}{2ma^2}\sum_{\mu=1,2}( 2\dl_{x,y} - \dl_{x,y+a\hat{\mu}}-\dl_{x,y-a\hat{\mu}} ),
    \label{Eq_Hxy}
\ee
where $a=L/N$ is the lattice distance, $m$ the particle mass; $x$ and $y$ are 2D vectors $(x_1,x_2)^T$ 
and $(y_1,y_2)^T$ with  $x_{1,2}$ and $y_{1,2} \in \{0,a,\dots,Na-a\}$;  the $\hat{\mu}$ are unit vectors,
 $\hat{1}=(1,0)^T$ and $\hat{2}=(0,1)^T$ and $\dl_{x,y}\equiv \dl_{x_1,y_1}\dl_{x_2,y_2}$ etc., with 
Kronecker $\dl_{x_{\mu},y_{\mu}}$.

For a box with impenetrable walls, the wave function at $x_{1,2}=-a$ and $aN$ must be zero. These
Dirichlet boundary conditions imply that Eq.~(\ref{Eq_Hxy}) must be modified at the boundary,
$\dl_{x,y+a\hat{\mu}}\ra \dl_{x,y}$ for $y_{\mu}=aN-a$ and $\dl_{x,y-a\hat{\mu}}\ra \dl_{x,y}$ for $y_{\mu}=0$;
With periodic boundary conditions, the Kronecker deltas at the boundaries are modified by 
identifying $y_{\mu}+a\hat{\mu}$ with $0$ if $y_{\mu}=aN-a$ 
and $y_{\mu}-a\hat{\mu}$  with $aN-1$ if $y_{\mu}=0$.

With Dirichlet boundary conditions the eigenvectors of $H$ are,
\be
   \psi^{(k)}_{x} = \sin((x_1 +a)k_1 \pi/(L+a))\sin((x_2+a) k_2 \pi/(L+a)),
\ee
with eigenvalues
\be
   E^{(k)} = [2 - \cos(k_1\pi/(N+1))-\cos(k_2\pi/(N+1))]/ma,
\ee
where $k=(k_1,k_2)^T$ and $k_1,k_2 \in \{1,\dots,N\}$. 
As in  refs.~\cite{ValentiniWestman05,ColinStruyve10} 
the initial wave function consists of a superposition of the $N_k^2$ lowest  energy eigenvectors multiplied 
with a random phase factor, such that the time dependent wave function is
\be
  \psi_x(t) = \sum_{k_1,k_2=1}^{N_k} e^{i\phi_k-itE^{(k)}}\psi_x^{(k)},   \label{Eq_Psi0}
\ee
with  $\phi_k$ a random phase between $0$ and $2\pi$. 
As an example, Figure~\ref{Fig_1} shows contour maps of the  probability distribution $\vert\psi(0)\vert^2$ 
in the 2D box for a low $\Delta P$ and high $\Delta P$ initial state ($N_k=4$ and $N_k=7$ respectively).

In most experiments below,  the wave function is a superposition of the 16 lowest-energy states 
($N_k = 4$); when exploring the impact of the momentum spread in the wave function, different
ranges of the wave number values will also be used, $N_k=3,\dots,8$.
Given limitations on compute capacity, it was not possible to explore systems in which both the lattice
distance in mass units, $ma$, would be very small and simultaneously the box dimension in mass units, $mL$, very large.
Particle masses will range from $mL=5$ to $40$, with most computations done at $mL=20$ and $mL=5$; lattices with different resolution will be employed with $N$ ranging from $15$ to $60$.

\begin{figure}[tth]
\begin{centering}
\includegraphics[height=7cm]{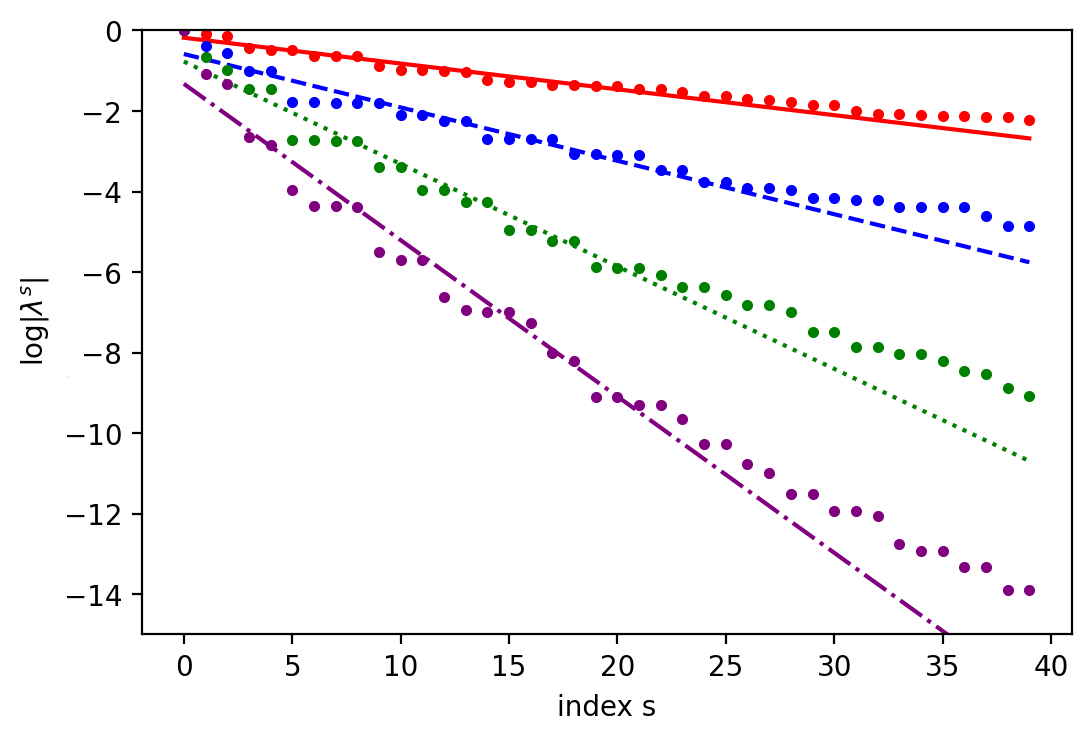}
  
\captionW{Eigenvalue spectra of $\Tt(t)$ evaluated on a system with Dirichlet boundary conditions, low
$L\Delta P = 8.55$ and $N=30$.  The  dots in the plot show $\log(\vert \lm^s\vert)$  
for $t/L =40, 120, 240$ and $360$, in order of increasing (negative) slope. The straight
lines are least-square fits to the first 25 data points.
\label{Fig_2}}
\end{centering}
\end{figure}

\begin{figure}[!thh]
\begin{centering}
\scalebox{1}{\includegraphics[height=3.85cm]{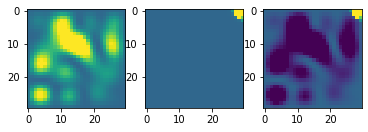}}\\[-2mm]
\scalebox{1}{\includegraphics[height=3.85cm]{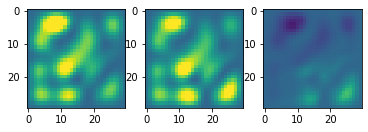}}\\[-2mm]
\scalebox{1}{\includegraphics[height=3.85cm]{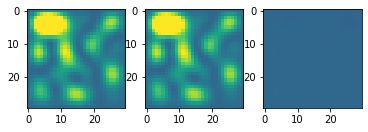}}\\[-2mm]
\end{centering}

\captionW{Density plots of the probability distribution $P(t)$ (left plots), the dominant right-eigenvalue $v^0(t)$ 
(middle plots) and the difference of $P(t)$ and $v^0(t)$ (right plots). The top, middle and bottom rows are at $t/L=0, 120$ and $360$ respectively.
\label{Fig_3}}
\end{figure}

For a given quantum system it is straightforward to compute $\Tt(t)$: Given the solution of the
Schr\"odinger equation shown in Eq.(\ref{Eq_Psi0}), the
time-dependent transition matrices $T(t)$ are  computed from Eqs.~(\ref{Eq_Tnm}) and (\ref{Eq_Tmm}),
after which $\Tt(t)$ follows from Eq.~(\ref{Eq_Tt}).
Both $\psi(t)$ and $\Tt(t)$ are evaluated at discrete times $t=k\eps$, with $\eps/L = 0.02/N$, such that the 
 time step size  $\eps/L$ and lattice distance $a/L$ both
scale as $1/N$.\footnote{To test the impact of  the time step size, a few computations were repeated with
$\eps/L=0.01/N$, which produced essentially identical results.}
 The number of time steps varies per simulation, ranging from $N_t=1000$ (for systems
with fast equilibration) to $50,000$ (for systems with  slow equilibration).

In the numerical evaluations of the evolving transition matrix it may happen 
that the constraint $\sum_{n\ne m}T_{nm} \le 1$ is violated for one or more columns $m$.
 When this happens (very rarely), 
the  diagonal element for the offending column is put to zero and the off-diagonals are normalized to sum up to 1. 
Since this could lead to inaccuracies, it is always checked that the
final probabilities computed using the Schr\"odinger equation (\ref{Eq_PSE}) and using the Markov process 
(\ref{Eq_PtFromT}) on the same initial probability are the same. In all cases the two final probabilities are very close:
$\sum_n \vert\psi^*_n(t)\psi_n(t) - (\Tt (t)P(0))_n\vert < 10^{-8}$ (with $P(0)=\vert\psi(0)\vert^2$).

\subsection{Eigenvalue Spectrum of the Cumulative Transition Matrix} \label{sect3sub2}
The objective of this subsection is to explore if the large-time eigenvalue spectrum of the cumulative transition matrix 
$\Tt(t)$ defined in Eq.~(\ref{Eq_Tt}) exhibits the desired dominance of the eigenvector with eigenvalue $1$
that is shown in Eq.~(\ref{Eq_Lm}).

Figure \ref{Fig_2} shows an example of the time-dependence of the eigenvalues of $\Tt(t)$, computed on a lattice
with $N=30$. The dots  represent $\log(\vert \lm^s\vert)$ for $s=0,1,\dots,40$ and 
the straight lines are least-square fits of the form $ b_0 + b_1 s$, using 
the first $25$ eigenvalues. From top to bottom, the spectra are for $t/L=40, 120, 240$ and $360$.
These results confirm first of all that the spectra only have a single eigenvalue $1$. Furthermore,
the absolute values of the eigenvalues decrease roughly exponentially with the eigenvalue
index (as in Eq.~(\ref{Eq_Lm}) with $w_s\approx s$). This suppression of eigenvalues grows with time and
the steepness of the lines in Fig.~\ref{Fig_2} should be a measure for the level of equilibration: the steeper the
slope, the closer the system is to quantum equilibrium.

The gradual approach to quantum equilibrium can also be seen from the evolution of the dominant right eigenvector.
As shown in section~\ref{sect2sub2}, this eigenvector $v^0(t)$ must  become equal to the quantum 
probability distribution $P(t)$ when $\Tt(t)$ obtains the form shown in Eq.~(\ref{Eq_Teq}). 
This gradual evolution can be seen in Fig.~\ref{Fig_3}, which shows
a density plot of the probability distribution $P(t)$ (left plots), the dominant right-eigenvector $v^0(t)$ (middle plots)
and their difference (right plots). The top row is at the initial state, $t/L=0$, the middle and bottom row are at 
$t/L=120$ and $360$ respectively.


\subsection{Quantum Equilibrium Times} \label{sect3sub3}
Having established that the eigenvalues $\lm^s$ with $s>0$ are (approximately) exponentially suppressed, 
\be
   \lm^s(t) \approx e^{-c(t)s},     \label{Eq_Lms}
\ee
this subsection will
explore the time-dependence of this suppression. This will lead to estimates of the equilibration time scale as a
function of the lattice distance. The dependence of the time scale on particle mass and momentum spread 
in the wave function will be discussed in the next subsection.

\begin{figure}[tth]
\begin{centering}
\includegraphics[height=7cm]{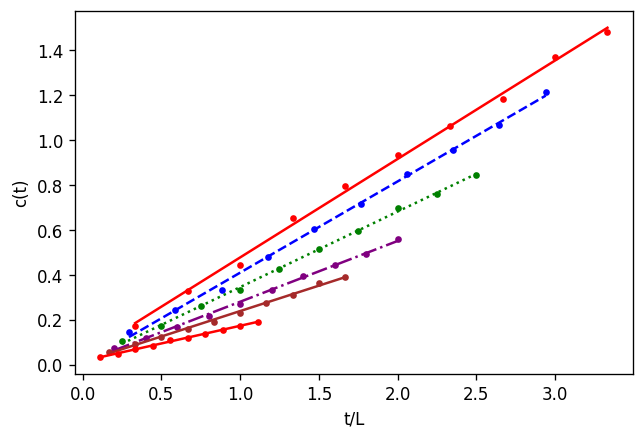}
  
\captionW{Time dependence of the slope-coefficients $c(t)$ in Eq.~(\ref{Eq_Lms}), for the closed box model with
 $mL=5$ and $L\Delta P = 8.55$, evaluated on lattices with different lattice distances $a=L/N$.
The straight lines are least-square fits to the data. From top to bottom the lines are for
$N=15, 17, 20, 25,  30$ and $45$.
\label{Fig_4}}
\end{centering}
\end{figure}

Figure \ref{Fig_4} shows the time dependence of the slope-coefficients $c(t)$ for the closed box system with $mL=5$
and $L\Delta P = 8.55$ evaluated on lattices with $N=15, 17, 20, 25, 30$ and $45$. 
As can be seen from the quality of the fit (which always have
$R^2$ values larger than $0.96$), the suppression of the non-dominant eigenvalue modes progresses to a 
good approximation linearly in time. Hence the slope-coefficient can be parameterized as
\be
   c(t) \approx c_0 + t/t^{eq},    \label{Eq_ct}
\ee
where the coefficient of $t$ is identified with the inverse of the equilibrium time scale $t^{eq}$. The 
intercept $c_0$ is small, typically less than $0.1$

\begin{figure}[!tht]
\begin{centering}
\includegraphics[height=7cm]{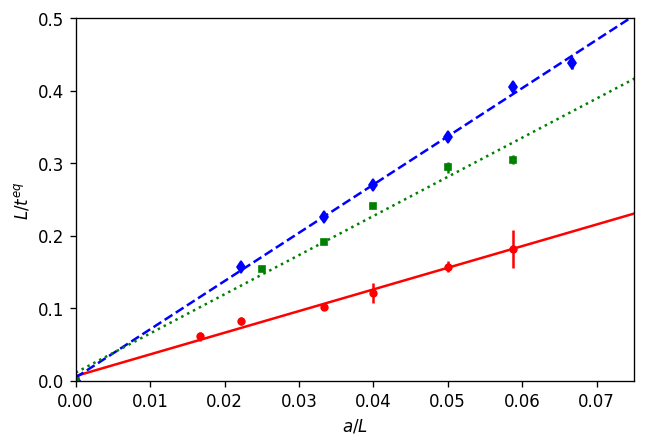}
  
\captionW{Discretization scale dependence of the equilibrium times. The plots
show $L/t^{eq}$ vs. $a/L$ for three  quantum systems: closed box with $mL=20$ (dots with
solid line), closed box with $mL=5$ (diamonds with dashed  line) and periodic box with $mL=20$ 
(squares with dotted line). The results of the fits are: 
$L/t^{eq} = 0.006(5)+3.0(1)(a/L)$, $L/t^{eq} = 0.004(5)+6.6(1)(a/L)$ and $L/t^{eq} = 0.01(2)+5.4(4)(a/L)$
respectively.
For the periodic box, the superposition of plane-wave energy states
has approximately the same $L\Delta P\approx 9$ as the wave functions in the closed box, which have 
$L\Delta P=8.55$. The lines are least-square fits through the data, including the point at the origin. 
\label{Fig_5}}
\end{centering}
\end{figure}

As was mentioned in 
the introduction,
the coarsening scale that is needed to demonstrate equilibration in the causal Bohm formulation 
appears to play the same role as the lattice distance in the stochastic Bell formulation. Consistency
between these two alternative formulations implies that the equilibrium time should diverge towards the
continuum limit $a\ra 0$. Adopting the expression for the equilibrium time, Eq.~(\ref{Eq_teq}) proposed by Valentini \cite{Valentini92,Valentini01,ValentiniWestman05} and equating the coarsening scale $\varepsilon$ with 
the lattice distance $a$, $t^{eq}$ is expected to grow linearly with $N = L/a$.
The results in Figure \ref{Fig_5} indeed show that $1/t^{eq}$ is roughly proportional to $1/N$. 

The data points in
this figure were obtained from linear fits of $c(t)$ vs.~$t$  in three 
different versions of the 2D quantum system: a closed box with Dirichlet boundary conditions and particle mass 
$mL=20$ (dots with solid line), a closed box with Dirichlet boundary conditions and particle mass $mL=5$ 
(diamonds with dashed  line), and an open box with periodic boundary condition with $mL=20$ 
(squares with dotted line). The wave function in the open box consists of a  superposition of 9 plane wave energy
eigenvectors, similarly randomized as the eigenvectors in the closed box, 
with roughly similar momentum spread. 

The straight lines in Figure \ref{Fig_5} are least square fits to the data points, where the error bars indicate the
standard error\footnote{As can be seen in Figure \ref{Fig_2}, the
exponential suppression of the eigenvalues of $\Tt(t)$ is not strictly linear in the eigenvalue index $s$: the
slope is not quite constant and there are clusters of near degenerate eigenvalues. This introduces additional
uncertainty in the fitted results for $c(t)$ at different values of $1/N$, which is not included in the size
of the error bars.} on the fitted coefficient of $t$ (in Eq.~(\ref{Eq_ct})).
The  slopes of the lines are slightly biased,
since also the point at the origin has been included in the fits. However, the trend in the data supports the expected
behavior that $L/t^{eq}\propto a/L$, and hence $t^{eq} \ra \infty$ for $a \ra 0$. 
This is also the case for the model  in which the box has
periodic boundary conditions. Here, the equilibrium time scales are actually systematically shorter than in the box with
hard wall boundaries (the dotted line in Figure \ref{Fig_4} lies above the solid line). This is somewhat surprising, since
the (rapid) establishment of quantum equilibrium in the causal Bohm approach has been linked to the level of 
chaotic behavior of the trajectories computed. One might intuitively think that  reflecting boundaries
would be important to promote chaotic behavior.

%

\subsection{Mass and Momentum Spread Dependence of Equilibrium Times}
\begin{figure}[tth]
\begin{centering}
\includegraphics[height=7cm]{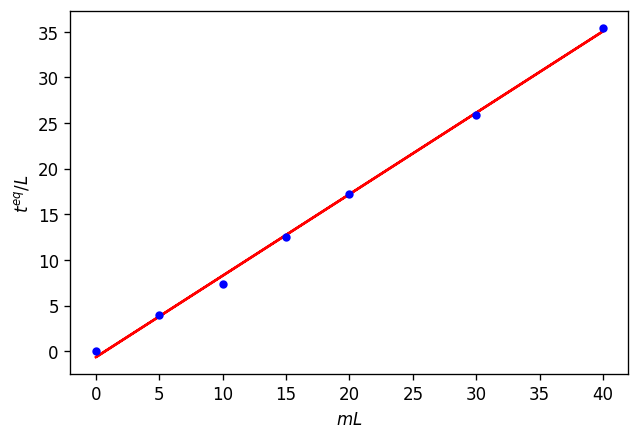}
  
\captionW{Mass dependence of the equilbrium times $t^{eq}/L$ for the closed box model with
 $N=30$ and $L\Delta P = 8.55$. The straight line, $t^{eq}=-0.6(3) + 0.89(2)mL$, 
is a least-squares fit of the data, including the point at the origin.
\label{Fig_6}}
\end{centering}
\end{figure}

As is expressed in the scaling relations (\ref{Eq_teq}) and (\ref{Eq_teq2}), the equilibrium time scale is
expected to grow with increasing particle mass and decrease with increasing momentum spread in the
system's wave function. Figure \ref{Fig_5} indeed shows that $L/t^{eq}$ values for $mL=20$ are smaller
than those for $mL=5$. To further explore the mass dependence of the equilibrium time, $t^{eq}$ is 
evaluated for a range of masses on the closed-box system with $L\Delta P=8.55$ and $N=30$.
The results  in Figure \ref{Fig_6} convincingly show that $t^{eq}/L$ grows linearly with $mL$,
as expected.

\begin{figure}[!tth]
\begin{centering}
\includegraphics[height=7cm]{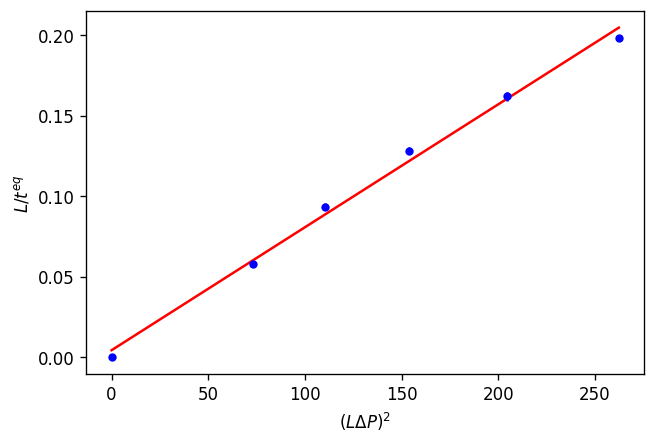}
  
\captionW{Momentum spread dependence of the equilbrium times $t^{eq}/L$ for the closed box model with
 $N=30$ and $mL=20$. The wave functions are composed of the $N_k^2$ lowest energy eigenvectors,
with $N_k=4,5,\dots,8$ ($L\Delta P = 8.55, 10.5, 12.4, 14.3$  and $16.2$). The straight line
, $L/t^{eq} = 0.004(4)+ 0.00076(3)(L\Delta P)^2$, is a least-squares fit of the data including the point at the origin.
\label{Fig_7}}
\end{centering}
\end{figure}

In analogy with thermal equilibration, the quantum equilibrium time is  expected to also depend on the 
momentum or energy spread in the system's wave function \cite{Valentini92,Valentini01}. 
This is confirmed by the results in Figure \ref{Fig_7}, which shows the $\Delta P$ dependence of $L/t^{eq}$.
The momentum spread $\Delta P$ is computed from the standard deviation of the
energy: $L\Delta P = L(2m\Delta E)^{1/2}$ with  $\Delta E = (\av{H^2}-\av{H}^2)^{1/2}$.
A least-squares fit $L/t^{eq}$ vs.~$(L\Delta P)^3$ (inspired by Eq.~(\ref{Eq_teq})) gives a relatively poor 
result ($R^2=0.94$); a linear fit of
$\log(L/t^{eq})$ vs. $\log(\Delta P)$ indicates a lower power of $L\Delta P$, $1.91(9)$ (and $2.0(1)$
 if the largest $\Delta P$ value is left out). This suggest to use a power 2 instead of 3, which leads to the 
result shown in Figure \ref{Fig_7}, where $R^2=0.99$.

Combining this result with that of the previous subsection, the scaling relation for the quantum equilibrium 
time using dimensionless parameters takes the form,
\be
   t^{eq}/L \propto  mL/(a/L)(L\Delta P)^2,
\ee
which in dimensionful parameters is the result  anticipated in Eq.~(\ref{Eq_teq2}) of the introduction,
\be
  t^{eq} \propto mL\hbar/a(\Delta P)^2.
\ee
The factor $m$ is easy to understand, because for a free particle the evolution matrix only
depends on the combination $\eps/m$, i.e., time scales are naturally proportional to the particle mass. 
Or, in the particle picture, the probability to move to a neighboring location (i.e.,
its typical velocity) is proportional to $\eps/am \propto 1/Lm$.

The appearance of a factor $L/a=N$ is not unreasonable
as one would expect that relaxation requires
the particles to travel  some system-specific distance, a mean free path lengths in a multi-particle system
or the distance between two low-probability regions in this single-particle example.
Since a particle will hop at most to a neighboring location,
$N$ sets the scale for the number of particle moves that are needed to cover such a distance.
Or, given the sparse nature of the individual transition matrices, which
only have non-zero off-diagonal elements for nearest neighbors (in at most one direction),
it is also clear that at least $N$ multiplications are needed to obtain a cumulative matrix in which all elements
are larger than zero.

Even though it seems reasonable to expect that an initial state with a large momentum or energy spread equilibrates
faster than one with a smaller spread, it is not true that initial states with zero momentum spread cannot equilibrate.
Obviously, an energy eigenstate that is real will never equilibrate, since the transition matrix for such a system
has no off-diagonal elements. As is the case in the causal Bohm formulation,
particles do not move when the system is in a real-valued energy eigenstate. However, on a space with periodic
boundary conditions, where energy eigenstates can be take as complex-valued plane waves, particles do move and
perhaps surprisingly, even a system with a single plane wave as initial state is found to equilibrate. This can happen
owing to the discrete time formulation in which the probability current (\ref{Eq_Jt}) is split over two time
steps. When expanding the transition probabilities in orders of the time step size $\eps$, one will find subleading
terms that allow transitions between neighboring locations (in the direction orthogonal to the direction the 
particle moves in), even though the wave function values at these locations evaluated at the same
time are equal. In the example of a single plane wave, the equilibrium time scale is found to 
increase with  decreasing time step size - as expected.  

As the above comments already suggest, it is not easy to find initial states that cannot support quantum equilibration,
i.e., states for which the transition matrix is non-primitive. 
Wave functions composed of random superpositions of energy states, but such that they possess 
a rotation or mirror symmetry that is preserved during time evolution, still 
lead to primitive transition matrices and  these systems readily equilibrate.
Besides real-valued energy eigenstates, it turns out that  initial (Gaussian) states that are sufficiently localized 
start out with a transition matrix that has multiple eigenvalues $1$. However, once the wave packet spreads out to
cover enough of the box volume, such that the transition matrix develops off-diagonal elements connecting
neighboring locations everywhere in the system, 
also here transition matrices become primitive, only have a single isolated eigenvalue $1$, 
 and also this system readily equilibrates.

\section{Discusion} \label{sect4}
The stochastic version of the de Broglie-Bohm formulation of quantum mechanics introduced by Bell,
can be rigorously defined on discrete space-time \cite{Vink22}. As was shown in section \ref{sect2},
the dBBB dynamics for the probability distribution of particle locations resembles a Markov process 
with  transition matrices $T(t)$ that are slowly varying in time.
This analogy with a Markov process, supported by proven properties  of the backward product
of such matrices \cite{Artzrouni91} suggests that the dBBB dynamics (for sufficiently small time step sizes)
will be weakly ergodic if the transition matrices are primitive (i.e., if they are such that all matrix elements
of $T^k(t)$ are positive for a sufficiently large power $k$). Weak ergodicity then implies that for large enough times, 
the cumulative transition matrix $\Tt(t)$ is dominated by a single eigenmode, as shown in Eqs.~(\ref{Eq_Teq}) 
and (\ref{Eq_wErg}),  which in turn implies that the system will relax to quantum equilibrium irrespective 
of the initial probabilities attributed to the particle locations in the ensemble. 

Direct calculations of the spectral decomposition of $\Tt(t)$ for
 a free particle in a 2D box confirm that this dynamical relaxation to quantum
equilibrium indeed happens for many variations of the system's wave function - in fact, it proves to be
difficult to find (non-static) wave functions for which equilibrium relaxation does not happen.
The example of a spreading Gaussian wave package, which starts to equilibrate once the package has sufficiently
spread out over the box volume, illustrates that a necessary condition for equilibration (i.e., for the system's
transition matrix to be primitive) is that the (non-static) wave function is non-zero everywhere, such that there 
is a non-zero transition probability for the particle to move, at every location in the system -- it may well be that this
condition is also sufficient.

The numerical results
furthermore confirm that the equilibrium time scale diverges when the lattice distance approaches zero.
This has to be the case, since in this limit the stochastic Bell dynamics must reproduce \cite{Vink93} the 
 causal Bohm dynamics for which quantum equilibration cannot  happen without introducing
coarse-graining \cite{Valentini92,Valentini01,ValentiniWestman05}. Since the quantum world in the discrete 
dBBB formulation is naturally discrete (and finite), such an additional coarse-grainig step, 
or interpretation, is not needed.

The computed values of $t^{eq}$  are sufficiently accurate to establish the scaling behavior of  equilibrium 
times shown in Eq.~(\ref{Eq_teq2}), which has the expected dependence on particle mass and lattice distance.
Unlike in Valentini's proposal shown in Eq.~(\ref{Eq_teq}), there is 
an additional dependence on the system size, with a correspondingly 
weaker dependence on the momentum spread.

Even though the simulations in this paper were performed on a very simple quantum system, the
proven properties of slowly evolving sequences of transition matrices,
 corroborated by these simulation results
 are sufficiently encouraging to expect that, for typical quantum systems of interest, the stochastic dBBB dynamics 
will, after a system-dependent relaxation time, reproduce the probability distributions computed with quantum 
mechanics irrespective of the details of the  initial particle distribution.
This dynamical relaxation to quantum equilibrium  elegantly counters one of the 
more persistent objections against Bohm-type interpretations of  quantum mechanics.

\subsection*{Acknowledgements}
I would like to thank Marc Artzrouni for helpful comments and discussion.

\end{document}